\documentclass[reprint,aps,prmaterials,amsmath,amssymb,superscriptaddress,floatfix]{revtex4-2}

\usepackage{eqnarray,amsmath}
\usepackage{lmodern}
\usepackage{graphicx}
\usepackage{xcolor}
\usepackage{soul}
\usepackage{booktabs}
\usepackage{braket}
\usepackage{csquotes} 

\usepackage{hyperref}
\hypersetup{colorlinks=true,allcolors=blue,pdfstartview={FitH}}
\usepackage[all]{hypcap}

\newcommand{\tuwieniap}{Institute of Applied Physics, TU Wien,  Vienna, Austria}
\newcommand{\tuwienmatchem}{Institute of Materials Chemistry, TU Wien, Vienna, Austria}
\newcommand{\zaragoza}{Instituto de Nanociencia y Materiales de Aragón (INMA), CSIC-Universidad de Zaragoza, Zaragoza, Spain}

\DeclareRobustCommand{\hematite}{\texorpdfstring{$\alpha\text{-Fe}_2\text{O}_3(1\overline{1}02)$}{a-Fe2O3(1-102)}}
\DeclareRobustCommand{\hematiteOneby}{\hematite{}-\woods{1}{1}{}}
\DeclareRobustCommand{\iridiumox}{\texorpdfstring{$\text{Ir}(100)$}{Ir(100)}-\woods{2}{1}{}O}
\DeclareRobustCommand{\ptrh}{\texorpdfstring{$\text{Pt}_{25}\text{Rh}_{75}(100)$}{Pt25Rh75}-\woods{3}{1}{}O}
\DeclareRobustCommand{\pt10x10}{\texorpdfstring{$\text{Pt}(111)$}{Pt(111)}-\woods{10}{10}{}Te}

\DeclareRobustCommand{\iv}{\texorpdfstring{$I(V)$}{I(V)}}
\DeclareRobustCommand{\leediv}{LEED~\iv{}}
\DeclareRobustCommand{\leedivdash}{LEED-\iv{}}
\DeclareRobustCommand{\deltas}{$\delta{}A_{i,\mathbf{g}}$}
\DeclareRobustCommand{\program}[1]{\textsc{#1}}
\DeclareRobustCommand{\tenserleed}{\program{TensErLEED}}
\DeclareRobustCommand{\python}{\program{Python}}
\DeclareRobustCommand{\calc}{\program{viperleed.calc}}
\DeclareRobustCommand{\vprjax}{\program{viperleed-jax}}
\DeclareRobustCommand{\ViPErLEED}{ViPErLEED}

\DeclareRobustCommand{\lmax}{$\ell_\text{max}$}

\DeclareRobustCommand{\Rfactor}{\texorpdfstring{$R$ factor}{R factor}}
\DeclareRobustCommand{\RfactorDash}{\texorpdfstring{$R$-factor}{R-factor}}
\DeclareRobustCommand{\PendryR}{Pendry's \Rfactor{}}

\DeclareRobustCommand{\woods}[3]{%
\texorpdfstring{\mbox{$(#1\hspace{0.1em}\times\hspace{0.1em}#2)$}}{(#1x#2)}%
\if\relax\detokenize{#3}\relax%
\relax%
\else%
\texorpdfstring{$R$#3$^\circ$}{R#3}%
\fi
}

\newlength{\pagefigure}
\newlength{\columnfigure}
\DeclareRobustCommand{\floatref}[3][]{%
\hyperref[#2]{#3}~\ref{#2}%
\if\relax\detokenize{#1}\relax%
\relax%
\else%
\hyperref[#2]{(#1)}%
\fi}

\DeclareRobustCommand{\Fig}[2][]{\floatref[#1]{#2}{Fig.}}

\DeclareRobustCommand{\Eq}[1]{Eq.~(\ref{#1})}

\DeclareRobustCommand{\SubFig}[2]{\Fig{#1}\hyperref[#1]{#2}}
\DeclareRobustCommand{\Section}[1]{\hyperref[#1]{Section}~\ref{#1}}
\DeclareRobustCommand{\RefInline}[1]{Ref.~\onlinecite{#1}}

\DeclareRobustCommand{\doiRef}[1]{\href{https://doi.org/#1}{doi:#1}}

\DeclareRobustCommand{\todoHidden}[1]{}

\DeclareRobustCommand{\SIName}{Supporting Information}
\DeclareRobustCommand{\SI}{\SIName{} \cite{supporting}} 
\DeclareRobustCommand{\SISection}[1]{Section~S#1 of the \SI{}}

\begin{document}

\title{Structural Optimization in Tensor LEED Using a Parameter Tree and $R$-Factor Gradients}

\date{\today}

\author{Alexander M. Imre}
\affiliation{\tuwieniap}
\author{Paul Haidegger}
\affiliation{\tuwieniap}
\author{Florian Kraushofer}
\affiliation{\tuwieniap}
\author{Ralf Wanzenböck}
\affiliation{\tuwienmatchem}
\author{Tobias Hable}
\affiliation{\tuwieniap}
\author{Sarah Tobisch}
\affiliation{\tuwieniap}
\author{Marie Kienzer}
\affiliation{\tuwieniap}
\author{Florian Buchner}
\affiliation{\tuwienmatchem}
\author{Jesús Carrete}
\affiliation{\zaragoza}
\author{Georg K. H. Madsen}
\affiliation{\tuwienmatchem}
\author{Michael Schmid}
\affiliation{\tuwieniap}
\author{Ulrike Diebold}
\affiliation{\tuwieniap}
\author{Michele Riva}
\email[Corresponding author: ]{riva@iap.tuwien.ac.at}
\affiliation{\tuwieniap}

\begin{abstract}
Quantitative low-energy electron diffraction [LEED $I(V)$] is a powerful method for surface-structure determination, based on a direct comparison of experimentally observed \iv{} data with computations for a structure model.
As the diffraction intensities $I$ are highly sensitive to subtle structural changes, local structure optimization is essential for assessing the validity of a structure model and finding the best-fit structure.
The calculation of diffraction intensities is well established, but the large number of evaluations required for reliable structural optimization renders it computationally demanding.
The computational effort is mitigated by the tensor-LEED approximation, which accelerates optimization by applying a perturbative treatment of small deviations from a reference structure.
Nevertheless, optimization of complex structures is a tedious process.

Here, the problem of surface-structure optimization is reformulated using a tree-based data structure, which helps to avoid redundant function evaluations. In the new tensor-LEED implementation presented in this work, intensities are computed on the fly, eliminating limitations of previous algorithms that are limited to precomputed values at a grid of search parameters. It also enables the use of state-of-the-art optimization algorithms.
Implemented in \python{} with the JAX library, the method provides access to gradients of the \Rfactor{} and supports execution on graphics processing units (GPUs). Based on these developments, the computing time can be reduced by more than an order of magnitude.
\end{abstract}

\maketitle


\section{Introduction\label{sec:intro}}

Low-energy electron diffraction (LEED) is a widely utilized technique in modern surface science due to its ability to quickly provide information on the structure of single-crystal surfaces.
A sharp diffraction pattern confirms the long-range order and reveals the periodicity and symmetry of the surface structure.
Beyond qualitative analysis, quantitative examination of diffraction intensities as a function of the incident electron energy, producing so-called \iv{} curves, offers deep insights into the atomic-scale arrangement of the surface, requiring only small investment in additional instrumentation.
\leediv{} combines surface sensitivity with high precision, making the method a valuable tool for the analysis of surface structures \cite{pendryLowEnergyElectron1974,vanhoveSurfaceCrystallographyLEED1979,vanhoveLowEnergyElectronDiffraction1986,
moritzSurfaceStructureDetermination2022,heinzElectronBasedMethods2013,woodruff60YearsSurface2024,
heldStructureDeterminationLowenergy2025,fausterSurfacePhysicsFundamentals2020,heinzElectronBasedMethods2013}.

Similarly to x-ray diffraction, LEED diffraction intensities encode information about atomic positions and vibration amplitudes on the picometer scale.
However, unlike x-rays, the electrons used in \leediv{}—with typical energies ranging from fifty to a few hundred electronvolts—have an inelastic mean free path of only a few atomic layers \cite{seahQuantitativeElectronSpectroscopy1979}.
As a result, the elastically scattered electrons that contribute to the diffraction intensities interact only with the uppermost atomic layers. In addition, elastic-scattering cross sections are large, leading to strong contributions of multiply scattered electrons to the diffraction signal and enhancing the sensitivity to small structural changes.
This unique combination of surface specificity and high sensitivity to interatomic distances makes \leediv{} the most successful method to verify or exclude structure models of surfaces \cite{heinzElectronBasedMethods2013,fausterSurfacePhysicsFundamentals2020,woodruff60YearsSurface2024,heldStructureDeterminationLowenergy2025}.

Unfortunately, like most diffraction techniques, \leediv{} suffers from the well-known phase problem, where only the intensities of the diffracted beams are accessible in the experiment.
The loss of the phase information and the contributions of multiple scattering make it impossible to directly derive the atomic positions from the measured intensities (with the exception of selected cases where holographic LEED or other \enquote{direct} methods can aid structure determination %
\cite{saldinThreedimensionalReconstructionHolographic1995,reuterLEEDholography1998,%
saldinSolutionMultipleScatteringInverse2002,
 changDirectLEED1999}).
Instead, a \leedivdash{} study requires an initial structure model (a result of a search using computational methods or a human \enquote{guess}), calculating the diffraction intensities for this model and refining the model by minimizing the disagreement between calculated and experimental \iv{} data. Thereby, qualitatively different structures can be evaluated and the parameters (such as the exact atomic positions) of a structure model can be determined.

Despite considerable progress in terms of usability and performance of \leediv{} calculations \cite{heldStructureDeterminationLowenergy2025,kraushoferViPErLEEDPackageCalculation2025}, the search for a best-fit structure is computationally expensive and often requires human intervention. With the currently most widely used programs, search is limited to a discrete grid and geometrical optimization is restricted to one direction at a time.  This makes \leediv{}  poorly suited for automated structure search, for instance, to assess the validity of structure models generated by machine-learning (ML) algorithms based on density-functional theory (DFT). The current work presents a new implementation for structural search by \leediv{}, which does not suffer from the limitations discussed above. Describing search parameters as a tree structure paves the way for the use of state-of-the-art search algorithms and computational optimization. Taking advantage of graphics processing units (GPUs) and automatic differentiation to calculate \RfactorDash{} gradients, the new implementation greatly reduces the computing time.




\subsection{\boldmath{}Basics of \leedivdash{} Calculations\label{sec:LEEDbasics}}

Theoretical modeling of \iv{} curves is a non-trivial task.
As mentioned in \Section{sec:intro}, the calculation must account for multiple scattering, also known as a \enquote{full-dynamic} calculation.
Techniques for solving this problem were first described by Pendry, Tong, van Hove, and collaborators in the 1970s, and many of their methods form the basis of \leedivdash{} calculations today \cite{pendryLowEnergyElectron1974,vanhoveSurfaceCrystallographyLEED1979,
vanhoveSurfaceCrystallographyLEED1979,vanhoveLowEnergyElectronDiffraction1986,moritzSurfaceStructureDetermination2022,
tongTheoryLowenergyElectron1975}.
Various codes have been developed over the years that can perform these calculations, each with their own strengths and target audiences
\cite{
  vanhoveAutomatedDeterminationComplex1993,
  blanco-reyFORTRAN90LowEnergyElectron2004,
  wanderNewModularLow2001,
  titteringtonCalculationLeedDiffracted1980,
  heldRealisticMolecularDistortions1996,
  blumFastLEEDIntensity2001,
  lachnittAQuaLEED,
  heldCLEEDPyCode}
.

Today's LEED codes are based on the muffin-tin approximation, with a constant potential $V_0$ (also referred to as inner potential) for the interstitial region and a spherical potential well for each atom \cite{pendryLowEnergyElectron1974,heldStructureDeterminationLowenergy2025}.
Based on this potential, scattering phase shifts can be calculated that describe a single scattering event of an incident electron wave with the scattering atoms 
\cite{rundgrenOptimizedSurfaceslabExcitedstate2003,rundgrenElasticElectronatomScattering2007,
rundgrenLowenergyElectronDiffraction2021,matererReliabilityDetailedLEED1995}.

Under that approximation, atoms are reduced to point scatterers that impart phase shifts onto the incoming electron wave.
The atoms are then grouped into crystal layers, and intra- and inter-layer scattering is calculated.
The scattering in between such layers is treated using plane waves, whereas the scattering within a layer uses a spherical-wave expansion.
This expansion is truncated at a maximum angular momentum quantum number \lmax{}.
The choice of \lmax{} is a trade-off between accuracy and computational cost, since the size of most intermediate tensors in the calculation scales with powers of \lmax{} up to $($\lmax{}$+1)^4$, and the number of floating-point operations scales with up to $($\lmax{}$+1)^6$.

Diffraction from the bulk crystal is modeled by infinitely repeating bulk-like layers at the bottom of the structure.
The final diffraction intensities are obtained by considering transmission, reflection, and attenuation for each crystal layer, and summing the possible diffraction paths \cite{tongTheoryLowenergyElectron1975,vanhoveSurfaceCrystallographyLEED1979}.

\subsection{The Optimization Problem\label{sec:optimization_intro}}

Because \leediv{} is extremely sensitive to structural details---down to picometer-scale changes in atomic positions---a single calculation of diffraction intensities is insufficient to assess the validity of a structure model.
Even for structures relaxed by DFT, the deviations between the calculated and actual coordinates may be large enough to yield a poor \Rfactor{} when compared to experiment; this may depend on the choice of the DFT functional \cite{feibelmanDFTRealWorld2010}.
Consequently, a valid structure model may be misjudged as incorrect. Thus, a local optimization against experimental data is required, even when only a qualitative yes/no decision is sought.

For structure optimization, the disagreement between calculated and experimental \iv{} curves is quantified by a single number, the \Rfactor{}, which should be minimized. The most commonly used $R$ factors are those of Zannazi and Jona \cite{zanazziReliabilityFactorSurface1977} and Pendry \cite{pendryReliabilityFactorsLEED1980}, with the latter being the preferred choice in most cases \cite{tearComparisonReliability$R$1981,vanhoveAutomatedDeterminationComplex1993,
spornAccuracyQuantitativeLEED1998}.
Recently, the authors of this article proposed a new \enquote{smooth} \Rfactor{} \cite{Imre2025PendryRfactor}, which addresses several shortcomings in its formulation.

The optimization problem can be formulated in terms of the model parameters: the atomic coordinates $\mathbf{r}_i$, the mean vibration amplitudes $v_i = \sqrt{\langle{}u_i^2\rangle_T}$, and, in case of substitutional alloys, compounds, or structures with (disordered) vacancies, the statistical occupancy values (concentrations) $c_i$ \cite{jonaRandomOccupationAdsorption1978}.
Here, the index $i$ runs over all scattering centers in the unit cell, that is, all geometric sites and all chemical species that may occupy them.
In addition, diffraction intensities also depend on nonstructural parameters (collectively denoted by $\mu$ in the following), such as the offset $V_{00}$ of the real part of the inner potential
\footnote{Further non-structural parameters are the angle of incidence of the electron beam or the imaginary part of the inner potential, $V_\mathrm{0i}$, which describes the attenuation by inelastic scattering. These parameters cannot be optimized with today's tensor-LEED method (see \Section{section:tensorleed_basics}) and are therefore not considered here. The methods for optimizing these parameters in \ViPErLEED{} \cite{kraushoferViPErLEEDPackageCalculation2025} are not modified by the current work.}.
For any vector $\xi = \{ \mathbf{r}_i, v_i, c_i, \mu \}$ of these input parameters (including all scattering centers $i$), the diffraction intensities $I_{\mathbf{g}}^{\mathrm{calc}}(\xi)$ can be calculated for all diffraction orders $\mathbf{g}$ and all electron energies. These calculated intensities are compared to the experimental data $I_{\mathbf{g}}^{\mathrm{exp}}$ through the associated \Rfactor{}, $R(I_{\mathbf{g}}^{\mathrm{calc}}(\xi); I_{\mathbf{g}}^{\mathrm{exp}})$.

This leads to a high-dimensional optimization problem with the \Rfactor{} as the objective function:
\begin{equation} \label{eq:optimization_problem}
    \min_{\xi} R(I_{\mathbf{g}}^{\mathrm{calc}}(\xi); I_{\mathbf{g}}^{\mathrm{exp}}).
\end{equation}
Any approach to tackle such a high-dimensional optimization problem inevitably requires a huge number of evaluations of the objective function, each involving the calculation of a full set of diffraction intensities.

While full-dynamic calculations give access to the \iv{} curves $I_{\mathbf{g}}^{\mathrm{calc}}(\xi)$ for arbitrary $\xi$, this calculation is computationally expensive
\footnote{
The computationally most expensive part of a full-dynamic calculation for large unit cells is the inversion of several $d \times d$ matrices, $d = N(\ell_{\mathrm{max}}+1)^2$ with $N$ being the number of scattering centers in the unit cell per layer and \lmax{} being the maximum angular-momentum quantum number used in the calculation \cite{tongTheoryLowenergyElectron1975,pendryLowEnergyElectron1974}.
Depending on the case and algorithm used, the computational effort for matrix inversions scales like $d^{2.38}$ to $d^3$ \cite{kraushoferViPErLEEDPackageCalculation2025}.
}.
A single reference calculation on a state-of-the-art workstation or supercomputer node requires on the order of minutes for anything but the smallest unit cells \cite{kraushoferViPErLEEDPackageCalculation2025}, and many hours for large structures \cite{moritzStructureDeterminationCoincidence2010,kisslingerSurfaceTelluridePhases2023}.
Although this cost is acceptable for single-shot calculations, it becomes prohibitive in a structure-optimization loop with a large parameter space.

\subsection{The Tensor-LEED Approximation\label{section:tensorleed_basics}}

To overcome the computational bottleneck of full-dynamic calculations, the so-called tensor-LEED approximation was introduced by Rous \textit{et al.} \cite{rousTensorLEEDTechnique1986,rousTheoryTensorLEED1989,rousTensorLEEDApproximationSurface1992}.
It treats small changes to an initial \enquote{reference} structure via a perturbative approach.
Compared to an optimization using only full-dynamic calculations, this substantially reduces the computational cost and therefore accelerates structure optimization by several orders of magnitude.
Perturbations to the reference structure may be geometric displacements of atoms, changes of vibration amplitudes, or even chemical substitutions \cite{lofflerInvestigationSurfaceAtom1994,dollChemicalSubstitutionSurface1993,heinzTensorLEEDGeometrical1996}.
If the reference structure is denoted as
\begin{equation} \label{eq:xi_ref}
  \xi^{\mathrm{ref}} = \{ \mathbf{r}_i^{\mathrm{ref}}, v_i^{\mathrm{ref}}, c_i^{\mathrm{ref}}, \mu^{\mathrm{ref}} \}
\end{equation}
and the perturbation as
\begin{equation} \label{eq:delta_xi}
  \delta \xi = \{\delta \mathbf{r}_i, \delta v_i, \delta c_i, \delta \mu \};
\end{equation}
then the perturbed structure is given by
\begin{equation}
  \xi = \xi^{\mathrm{ref}} + \delta \xi.
\end{equation}
The intensities of the diffracted beams are then given by
\begin{equation}
  I_{\mathbf{g}} = |A_{\mathbf{g}}^{\mathrm{ref}} + \delta A_{\mathbf{g}}(\delta \xi) |^2,
\end{equation}
where the amplitude changes $\delta A_{\mathbf{g}}$ (sometimes also called \enquote{delta amplitudes}) are a function of $\delta \xi$.
In tensor LEED, the amplitude changes for any beam $\mathbf{g}$ are approximated by a linear combination of the amplitude changes for each scattering center $i$ in the unit cell \footnote{%
    The original formulation allows different levels of sophistication for modeling the effects of geometric displacements, with correspondingly different computational cost. \citet{rousTensorLEEDTechnique1986} suggested a middle-ground approach in which individual atomic displacements affect the atomic $t$ matrices, while mixed terms involving multiple atoms are neglected. 
    },
\begin{equation} \label{eq:superposition_of_deltas}
  \delta A_\mathbf{g} = \sum_i \delta A_{i,\mathbf{g}}.
\end{equation}
Nonstructural parameters $\mu$ can affect both the individual $\delta A_{\mathbf{g}}$ and the overall \RfactorDash{} calculation.

Tensor LEED is a first-order perturbation theory with respect to the influence of the scattering matrices $t_i$ on the amplitudes $\delta A_{\mathbf{g}}$. Nevertheless, the amplitude changes \deltas{} are highly nonlinear functions of the atomic displacements. This nonlinearity is illustrated in \SISection{1}.
Consequently, a simple linearization of the diffraction amplitudes (or intensities) is not possible, and the optimization problem posed in \Section{sec:optimization_intro} remains nontrivial.

\subsection{ViPErLEED}

The current work is based on \ViPErLEED{}, an integrated package for \leedivdash{} experiments and simulations \cite{kraushoferViPErLEEDPackageCalculation2025,schmidViPErLEEDPackageII2025,
dorrViPErLEEDPackageIII2025,viperleeddevelopersViPErLEEDDocumentation}.
Its \calc{} \python{} package provides calculations of diffraction intensities, structure optimization, and comparison of calculated and experimental \iv{} curves.
Internally, \calc{} uses the Erlangen tensor-LEED package (\tenserleed{}) \cite{blumFastLEEDIntensity2001} for most calculations
\footnote{The \ViPErLEED{} source code is available on GitHub under \url{https://github.com/viperleed/viperleed}. \tenserleed{}~2.0 is available under \url{https://github.com/viperleed/viperleed-tensorleed}. For more information on the usage and installation, see \url{https://www.viperleed.org} \cite{viperleeddevelopersViPErLEEDDocumentation}}.

\subsection{Optimization in \tenserleed{} and \calc{}}
\label{sec:optimization_in_tenserleed}

In \tenserleed{}, for computational efficiency, the parameter space is sampled on a user-defined grid \cite{blumFastLEEDIntensity2001,kottckeNewApproachAutomated1997}.
For every atom, the user can specify a set of geometric displacements, vibration amplitudes, and occupancies to be considered during the optimization.
Based on the user input, the code first calculates the amplitude changes \deltas{} for all scatterers $i$ and requested grid points.
During the minimization, for each configuration sampled, $\delta A_{\mathbf{g}}$ is calculated as a linear combination of the \deltas{} [see \Eq{eq:superposition_of_deltas}] and the \Rfactor{} is calculated by comparing the calculated diffraction intensities to the experimental data.
The parameter combination that achieves the minimum \Rfactor{} is referred to as the best-fit structure.

While the method used in \tenserleed{} is effective, it also has several shortcomings that lead to inefficiencies.
The calculation of \deltas{} is strictly separated from the optimization step, and the optimization itself is grid-based.
The step width---and thus the achievable precision of the optimized structure---must be fixed before the optimization loop begins.
In addition, atoms can only be displaced along a one-dimensional trajectory, which restricts the search space and leaves many configurations inaccessible.
Thus, an iterative approach is required to optimize atomic positions in three-dimensional (3D) space.

\calc{} includes some provisions to work around these issues as described in \RefInline{kraushoferViPErLEEDPackageCalculation2025}, although it cannot completely circumvent them.
The structure optimization remains the bottleneck of the calculation, especially for large and complex surface structures, and often takes more than 90\% of the total calculation time.

Additionally, the structure optimization is often the most problematic step in a \leedivdash{} study from a user perspective.
An inexperienced user may select parameters (such as the step sizes) with which the optimization algorithm can become trapped at a local minimum.
The grid-based nature of the optimization and the limitation to one-dimensional atomic displacements make it difficult to set up the optimization efficiently.

\section{Tensor LEED Using Python, JAX, and Automatic Differentiation}

Despite the sophistication of the $R$-factor--minimiza\-tion algorithm \cite{kottckeNewApproachAutomated1997} used in \tenserleed{}, its improvements in \ViPErLEED{}, and further efforts to develop optimization methods tailored to \leedivdash{} \cite{cowellUnconstrainedOptimisationSurface1987,kleinleNovelProcedureFast1989,kleinleEfficientMethodLEED1990,rousDirectedSearchMethods1990,overOptimizationMethodsTheir1992,rousGlobalApproachSearch1993,vanhoveAutomatedDeterminationComplex1993,dollGlobalOptimizationLEED1996,nascimentoDifferentialEvolutionGlobal2015}, existing implementations often remain tightly coupled to specific systems or surface geometries.
While effective in context, this limits their generality.
Here, a more abstract formulation of \leedivdash{} structure optimization is introduced to improve transferability across systems.


\subsection{High-Dimensional Optimization and Automatic Differentiation}

As mentioned in \Section{sec:optimization_intro}, structure optimization in \leediv{} is a high-dimensional numerical optimization problem.
As such, it can (and should) be tackled by modern numerical optimization algorithms that have been developed over the past decades.
Advances in fields such as machine learning have led to the development of highly efficient and robust optimization algorithms that can be applied to a wide range of problems \cite{nocedalNumericalOptimization2006,bonnasNumericalOptimization2006,kochenderferAlgorithmsOptimization2019}.

Today, optimization can also make use of automatic-differentiation libraries that provide an efficient calculation of gradients for a wide range of functions.
While this development has been primarily driven by the machine-learning community and the demand for efficient back-propagation algorithms, the applications of automatic differentiation are not limited to the field 
\cite{margossianReviewAutomaticDifferentiation2019,baydinautodiff2017,grotschel2012reverseDifferentiation}.
Several machine-learning frameworks exist that support automatic differentiation and highly parallelized execution on graphics-processing units (GPUs) such as \program{Tensorflow} \cite{tensorflow2015-whitepaper}, \program{Pytorch} \cite{anselPyTorch2Faster2024}, and JAX \cite{jax2018github}.
The latter is used in the tensor-LEED implementation of this work.

\subsection{Application to Tensor LEED}

A generalized framework for structure optimization within \leediv{} and \calc{} is presented, based on a reworked tensor-LEED implementation with four key improvements:
\begin{enumerate}
  \item Automatic interpretation and dimensionality reduction of the parameter space using symmetry and user-defined constraints (\Section{sec:parameter_space} and Sections~S6--S9 of the \SI{}). 
  This approach enables all parameter changes (i.e., perturbations) to be represented by a single, normalized, real-valued vector.
  \item Exploitation of symmetry and composed transformations to minimize redundant evaluations of functions (e.g., spherical harmonics) at each scattering center. This substantially reduces computational cost and is especially valuable for complex systems.
  \item Implementation in \python{} with the JAX library \cite{jax2018github}, which supports GPU execution and provides gradients of the \RfactorDash{} hypersurface by automatic differentiation (Sections~\ref{sec:implementation} and \ref{sec:search_algorithms}). This enables the use of gradient-based optimizers.
  \item Integration of these improvements to enable the application of established algorithms to the optimization problem. The task is separated into an exploratory (i.e., global) search for the best-fit structure followed by a local refinement (Sections~\ref{sec:search_algorithms} and \ref{sec:benchmarks}).
\end{enumerate}

This new implementation, named \vprjax{}, takes a different approach than \tenserleed{}:
The calculation of the amplitude changes $\delta A_{\mathbf{g}}$ is incorporated into the structure-optimization step and done on the fly as required.
This allows calculating the \Rfactor{} at any point in the parameter space, and removes the need for a grid-based method.
As discussed in Sections \ref{sec:search_algorithms} and \ref{sec:benchmarks}, while this is computationally more expensive per configuration, it improves the performance of the search, leading to an altogether faster convergence.

\section{Encoding the Parameter Space}
\label{sec:parameter_space}

The parameter space for a tensor-LEED calculation is usually high-dimensional.
While the full set of mathematically possible perturbations is given by the combination of 3D geometric displacements $\delta\mathbf{r}_i$, changes of vibration amplitudes $\delta{}v_i$, and fractional occupancy values $\delta{}c_i$ for all scattering centers, plus nonstructural parameters $\delta \mu$, the effective parameter space is much smaller.
Symmetry (which is detected automatically in \calc, but can be overriden by the user) restricts the set of valid perturbations.
Furthermore, users commonly impose additional constraints on the perturbations, such as linked vibration amplitudes or concerted atomic displacements, to conform with physical intuition or avoid overfitting.
Even if these constraints are not directly required by symmetry, they may be necessary to ensure that the optimization problem is well posed.

\subsection{Linear Transformations of Parameters}
\label{sec:linear_transformations}

A constraint on the parameters $\xi$ reduces the number of degrees of freedom of the optimization problem.
If one considers, for example, a symmetry constraint that links two or more symmetry-equivalent atoms, the number of free geometric parameters of this group of atoms is reduced to a total of (at most) three.
The displacements $\delta \mathbf{r}_{a,b}$ of two symmetry-equivalent atoms $a$ and $b$ are linked by a linear symmetry operation $\mathbf{M}$ as
\begin{equation} \label{eq:symmetry_operations}
  \delta \mathbf{r}_b = \mathbf{M} \delta \mathbf{r}_a.
\end{equation}
Equivalently, any such symmetry-linked displacement $\delta \mathbf{r}_i$ may be described by a single parameter vector $\mathbf{p}_k$, such that
\begin{equation}
    \delta \mathbf{r}_i = X_{i , k}^{\mathrm{sym}} \mathbf{p}_k,
\end{equation}
where the transformations $X_{i , k}^{\mathrm{sym}}$ encode the symmetry information of the system.

A user may further choose to constrain the system, for instance, by linking out-of-plane displacements of non-symmetry-equivalent atoms in a crystal layer.
In this case, the displacements of the atoms can be described by a second linear transformation $X_{j , k}^{\mathrm{link}}$ acting on an even smaller set of parameters $\mathbf{t}_k$ as given by
\begin{equation}
  \delta \mathbf{r}_i = X_{i , j}^{\mathrm{sym}} X_{j , k}^{\mathrm{link}} \mathbf{t}_k.
\end{equation}

In both cases, the constraints reduce the number of degrees of freedom of the system, and allow determining the full set of displacements $\delta \mathbf{r}_i$ ($3n$ values for $n$ linked atoms) through a linear transformation applied to $\mathbf{t}_k$ (a maximum of three values).

It should be noted that there are also nonlinear constraints, which cannot be expressed as linear transformations. \SISection{10} shows that the occupancy parameters (concentrations) are such a case and how this can be handled in the current framework.

\subsection{Irreducible Parameter Space\label{sec:irreducible}}

The full set of parameters, $\xi = \{ \mathbf{r}_i, v_i, c_i, \mu \}$ \cite{post_processing_footnote}, belongs to a vector space $\Xi$.
Constraints reduce the degrees of freedom of the system and can be expressed as (linear) transformations acting on the parameters. This allows the full parameter set $\xi$ to be uniquely determined by a reduced set $\xi^{\mathrm{red}}$ from a subspace $\Xi^{\mathrm{red}}$, with
\begin{equation}
\dim(\Xi^{\mathrm{red}}) \leq \dim(\Xi).
\end{equation}

The mapping $f^{\mathrm{red}}$,
\begin{equation}
f^{\mathrm{red}} : \Xi^{\mathrm{red}} \rightarrow \Xi,
\end{equation}
is injective, meaning that no two distinct reduced parameter sets $\xi^{\mathrm{red}}$ map to the same full set $\xi$.

By applying all constraints, a final set of \enquote{irreducible} parameters, $\tilde{\xi}$, is obtained. These parameters belong to the irreducible subspace $\tilde{\Xi}$, where
\begin{equation}
  \tilde{\xi} \in \tilde{\Xi}, \quad \dim(\tilde{\Xi}) \leq \dim(\Xi).
\end{equation}

Consequently, the vector $\xi$ of full parameters, can be uniquely determined from the irreducible vector $\tilde{\xi}$ through an injective mapping
\begin{equation}\label{eq:mapping_function}
  f : \tilde{\Xi} \rightarrow \Xi, \quad \tilde{\xi} \mapsto \xi = f(\tilde{\xi}).
\end{equation}

The mapping $f$ encodes all constraints on the parameters $\xi$, and converts any $\tilde{\xi}$ to a set of input values for the tensor-LEED calculation \cite{post_processing_footnote}.
In other words, $f(\tilde{\Xi})$ spans the constrained parameter space within $\Xi$, covering all valid configurations of $\xi$.
The optimization algorithm works in the irreducible parameter space. The mapping $f$ translates each point $\tilde{\xi}$ probed by the optimization algorithm to an input for the tensor-LEED calculation.

\subsection{Tree-like Representation of the Parameter Space}
\label{sec:parameter_tree_math}

\begin{figure*} \label{fig:full_tree}
  \includegraphics[width=17.9cm]{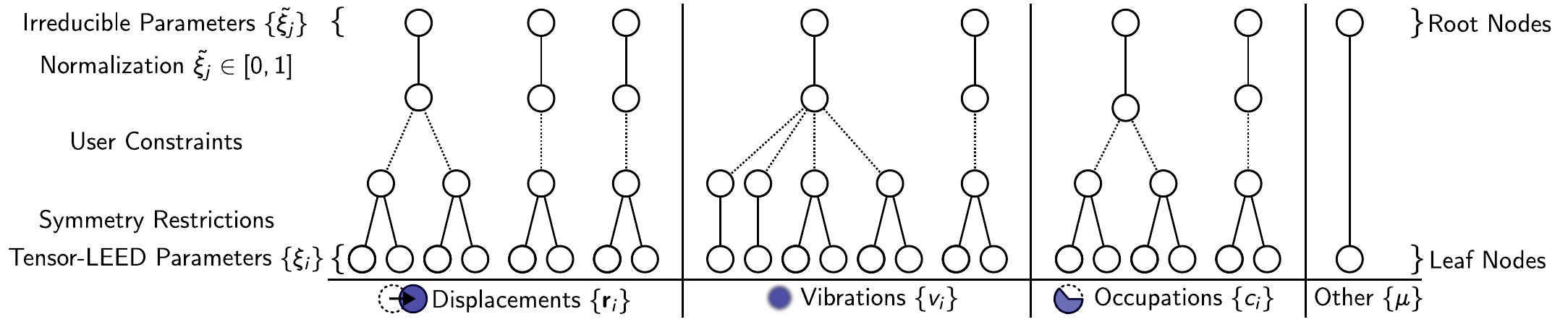}
  \caption{Schematic representation of the parameter tree. The input parameters of the tensor-LEED calculation are $ \xi = \{\xi_i\} = \{ \mathbf{r}_i, v_i, c_i, \mu \}$ \cite{post_processing_footnote}. They are linked to the vector of irreducible parameters $\tilde{\xi} = \{\tilde{\xi}_j\}$ via system and user constraints. Dashed lines represent links due to user constraints, which depend on user input and may be changed. Solid lines represent links which are automatically created and do not depend on user inputs (symmetry restrictions and normalizations).}
\end{figure*}

Borrowing concepts from graph theory, the application of constraints can be represented using a tree-like data structure, similar to computational graphs in automatic differentiation and machine learning \cite{baydinautodiff2017}.
Value propagation in this \enquote{parameter tree} helps to reduce the number of function evaluations needed to calculate the \Rfactor{} and its gradient.

Nodes in the tree (\Fig{fig:full_tree}) represent scalar- or vector-valued parameters, and edges correspond to linear or affine transformations between them.
The leaf nodes contain the full set of parameters $\xi$ of the tensor-LEED calculation \cite{post_processing_footnote}, while the root nodes form the irreducible set of free parameters $\tilde{\xi}$.
\Fig{fig:full_tree} highlights the hierarchical structure of the parameter space and the mapping between full and irreducible parameter sets via multiple layers of constraints.

Each non-leaf, non-root node has a single parent and one or more children.
The values of each child node are obtained by applying the transformation $X_e$ associated with the edge $e$ to the value of the parent node.
Constraints are naturally applied in a bottom-up manner: starting from the leaves and narrowing the parameter space toward the roots.
Although the order of constraints does not affect the uniqueness of the mapping between full and irreducible parameters, applying symmetry constraints first is advantageous, as these are the most fundamental ones and make the construction of the tree more efficient.

All parameters have user-specified bounds (e.g., a range for displacements).
These bounds may reflect physical limitations, since the tensor-LEED approximation is only valid for small perturbations, and vibration amplitudes must be positive.
For optimization, all parameter values are normalized such that the components of the irreducible parameters $\tilde{\xi_j}$ lie in the $[0,1]$ interval, as indicated in \Fig{fig:full_tree}.
Details of the normalization are given in \SISection{9}.

\subsection{Properties of the Parameter Tree\label{section:tree_properties}}

Sections~\ref{sec:dependent_quantities}--\ref{sec:bounds_and_normalization} describe how the tree-like representation of the parameter space can be used and leveraged.
To facilitate this discussion, some properties of the tree and the transformations applied to the parameters are introduced here.

\subsubsection{Linearity}

Each edge in the tree, connecting a parent node $n$ to a child node $n'$, is assumed to represents an affine transformation $X_{n', n}$ of the form
\begin{equation}
  X_{n', n}(p) = M_{n', n} p + b_{n', n}, \label{eq:affine}
\end{equation}
where $M_{n', n}$ is the matrix of a linear transformation acting on the (scalar or vectorial) parameter $p$, and $b_{n', n}$ a translation.
Except for normalization to the $[0,1]$ range of the irreducible parameters $\tilde\xi$, the transformations are linear mappings, without a translation (i.e., $b_{n', n} = 0$ in \Eq{eq:affine}; see \SISection{8}).
Affine transformations are efficiently implemented using matrix--vector multiplication and addition.
Since the tree is constructed only once per optimization run, its computational overhead is negligible.

\subsubsection{Composability and the Mapping Function\label{sec:composability}}

The transformation $X_{n_0,n_k}$ between any two nodes $n_0$ and $n_k$ is obtained by composing the transformations along the path connecting them:
\begin{equation}
  X_{n_0,n_k} = X_{n_0, n_1} \circ X_{n_1, n_2} \circ  \cdots \circ X_{n_{k-1}, n_k},
\end{equation}
where $n_1, n_2, \dots, n_{k-1}$ are the intermediate nodes along the path.

Thus, the mapping $f$ [\Eq{eq:mapping_function}] from irreducible parameters $\tilde{\xi}$ to the full parameters set $\xi$ is obtained by composing the transformations along the path connecting the corresponding nodes:
\begin{equation}
  \xi_i = f(\tilde{\xi_j}) = X_{i , j}(\tilde{\xi_j}).
\end{equation}

\subsubsection{Invertibility\label{section:invertability}}

For some calculations (dependent quantities, see \Section{sec:dependent_quantities}, propagation of bounds, see \SISection{9}), 
it is necessary to determine the transformation from a child node $n'$ to its parent $n$, $X_{n, n'}$, which is the inverse of $X_{n', n}$. Affine transformations are not always invertible; for example, a symmetry-violating displacement cannot be expressed in terms of irreducible parameters. However, as mentioned in \Section{sec:irreducible}, all transformations considered here are injective.
A non-invertible case cannot occur because the values at the child node $n'$ always lie in the image of $X_{n', n}$. For the injective linear transformations considered here (see also \SISection{8}), the inverse exists and is given by the Moore--Penrose pseudoinverse
\footnote{For an injective linear transformation with associated matrix $A$, a left pseudoinverse $A^+$ exists such that
\begin{equation}
  A^+ A = \mathbb{I},
\end{equation}
where $\mathbb{I}$ is the identity matrix. The pseudoinverse is
\begin{equation} \label{eq:pseudo_inverse}
  A^+ = (A^* A)^{-1} A^*,
\end{equation}
with $A^*$ the Hermitian conjugate of $A$. For real-valued transformation matrices, $A^*$ is replaced by the transpose $A^\top$.

Thus, the inverse of an injective linear transformation $X_{n , m}$ in the parameter tree is
\begin{equation}
  X_{m, n} = M_{n , m}^{+} = (M_{n , m}^\top M_{n , m})^{-1} M_{n , m}^\top,
\end{equation}
where $M_{n , m}$ is the (real-valued) matrix representation of the linear transformation $X_{n , m}$.
}.

\subsection{Tree Traversal and Dependent Quantities\label{sec:dependent_quantities}}

Since the transformations corresponding to the edges of the parameter tree are invertible, the transformation between two nodes with the same root can be computed.
For instance, within a group of linked atoms the displacement $\delta \mathbf{r}_i$ of atom $i$ can be expressed in terms of the displacement $\delta \mathbf{r}_{i'}$ of a reference atom $i'$ as
\begin{equation}
  \delta \mathbf{r}_i = X_{i , i'} \delta \mathbf{r}_{i'}.
\end{equation}
Here, $X_{i , i'} = X_{i , a} \circ X_{a, i'}$ denotes the composed transformation between nodes $i$ and $i'$, obtained by traversing the tree upward from the reference atom $i'$ to a common ancestor node $a$, and then downward to atom $i$.

Although seemingly trivial, this relation is central to a highly efficient implementation of tensor LEED, since the change in the diffraction amplitudes is obtained as a superposition of atomic contributions $\delta A_{i,\mathbf{g}}$ \cite{rousTensorLEEDTechnique1986,rousTheoryTensorLEED1989}, as given in \Eq{eq:superposition_of_deltas}.
The evaluation of $\delta A_{i,\mathbf{g}}$ for each vector of parameter variations $\delta \xi_i$ dominates the computational cost, as it requires many evaluations of computationally expensive functions (spherical Bessel functions, spherical harmonics) for each scattering center. Since displacements of linked atoms can be obtained from those of a reference atom via composed transformations, only a reduced set of atomic contributions $\delta A_{i,\mathbf{g}}$ needs explicit computation. This avoids redundant evaluations and yields substantial efficiency gains. The strategy is referred to here as \enquote{dependent quantities} and is detailed in \SISection{5}.

\subsection{Atoms as Leaf Nodes\label{sec:atoms_as_leaf_nodes}}

Sections S7 and S8 of the \SI{}  discuss the practical construction of  the parameter tree for a tensor-LEED calculation.
The tree is built from the leaves upward to the root, with symmetry relations and user-defined constraints enforced along the way.
This construction is not only suited for structural optimization, but also provides a systematic means of verifying the consistency of user-imposed constraints.

As introduced in \Section{sec:optimization_intro}, the input parameters for a tensor-LEED calculation are given by $\xi = \{ \mathbf{r}_i, v_i, c_i, \mu \}$ \cite{post_processing_footnote}, where $i$ indexes all scattering centers, encompassing the geometric sites and, in the case of mixed occupancy, all chemical species that may occupy them.
A natural choice is therefore to assign leaf nodes to each scattering center and parameter type (geometric, vibrational, occupancy).
To ensure consistent handling, the nonstructural parameters $\mu$ are included as additional leaf nodes, as illustrated in \Fig{fig:full_tree}.

\subsection{Bounds, Implicit Constraints, and Normalization\label{sec:bounds_and_normalization}}

As mentioned in \Section{sec:irreducible}, the optimization problem of \Eq{eq:optimization_problem} is subject to bounds on the parameters $\xi_i$,
\begin{equation}
  \xi_i^{\mathrm{lower}} \le \xi_i \le \xi_i^{\mathrm{upper}}.
\end{equation}


Parameters can also be completely fixed (e.g., restricting an atom to displacements normal to the surface plane even in the absence of symmetry constraints that fix the $x$ and $y$ coordinates).  
Such fixed parameters act as implicit constraints, reducing both the number of degrees of freedom and the dimensionality of the vector $\tilde{\xi}$ of irreducible parameters.

Both bounds and implicit constraints can be incorporated directly into the mapping function~$f$ of \Eq{eq:mapping_function}.  
By construction, $f$ maps the irreducible parameters $\tilde{\xi}_j \in [0,1]$ onto the full parameters $\xi_i \in [\xi_i^{\mathrm{lower}}, \xi_i^{\mathrm{upper}}]$.  
This formulation ensures that any optimization algorithm operates on normalized parameter vectors $\tilde\xi$ with equally weighted components, which can improve convergence.  
Implementation details for this normalization can be found in \SISection{9} \cite{zieglerLecturesPolytopes1995}.

\section{Implementation in \textsc{viperleed-jax} \label{sec:implementation}}

The new developments for tensor LEED described above are realized in \vprjax{} \cite{imreViperleedjax}.
Written entirely in \python{}, it is distributed as open-source software under the GNU General Public License version 3 (or later) \cite{gnuv3}. This implementation can be chosen to replace the \tenserleed{}-based structure optimization of  \calc{} \cite{kraushoferViPErLEEDPackageCalculation2025}, which is released under the same license.

\subsection{Parameter Tree Implementation}

The code is designed to be modular and easily extensible, enabling adaptation to specific applications.
In particular, the tree-like parameter representation is implemented separately from the optimization routines and includes several layers of abstraction.
It depends only on the \program{Numpy} \cite{harrisArrayProgrammingNumPy2020}, \program{SciPy} \cite{2020SciPy-NMeth}, and \program{Anytree} \cite{AnytreePythonTree} packages.
This separation facilitates extension to problems beyond \leediv{} or the incorporation of additional types of constraints or nonstructural parameters.

\subsection{Tensor-LEED Implementation in JAX}

The new tensor-LEED code in \vprjax{} makes use of the output of the full-dynamic calculation (the \enquote{reference calculation}) of \calc{} (which internally uses the \tenserleed{} code) and keeps compatibility with the data formats of \calc{}.  The calculation of the \iv{} curves by tensor LEED, the calculation of the \Rfactor{} and the search have been entirely re-implemented in \python{} and JAX \cite{jax2018github} to enable end-to-end differentiability and just-in-time compilation for GPU-accelerated execution.
Most required mathematical functions are available in JAX's standard libraries, with additional functionality supplied by the \program{NeuralIL} package \cite{montes-camposDifferentiableNeuralNetworkForce2022} for spherical Bessel functions and the \program{Interpax} package \cite{Conlin_interpax} for spline interpolations.
The implementation is designed as a plugin for \calc{}, serving as an optional substitute for the \tenserleed{}-based structure search.

To support automatic differentiation and efficient tensor contractions, several computational strategies and best practices are employed. Selected techniques are discussed in \SISection{4}, including the use of precalculated Gaunt coefficients, evaluated with \program{SymPy} \cite{10.7717/peerj-cs.103} and stored according to a scheme inspired by \citet{raschEfficientStorageScheme2004}.

\section{Search Algorithms}
\label{sec:search_algorithms}

The \RfactorDash{} hypersurface, $R({\tilde{\xi}})$, depends on the surface structure and the chosen \Rfactor{}, and is a highly nonlinear function of the parameters $\tilde{\xi}$ with numerous local minima \cite{vanhoveLowEnergyElectronDiffraction1986,kottckeNewApproachAutomated1997}.
Finding the global minimum becomes increasingly difficult as the number of parameters increases.

No single algorithm is universally optimal, and the choice of method depends on the specific problem \cite{wolpertNoFreeLunch1997}.
The approach presented in \Section{sec:parameter_space} reduces the structure optimization in \leediv{} to a standard bounded numerical optimization problem. Thus, the \Rfactor{} can be treated as a black-box objective function that takes the normalized irreducible parameters $\tilde{\xi}$ as input and returns a single scalar value, optionally with a gradient. This makes it easy to try different algorithms among the large number of optimization methods developed for numerical optimization. In the present work, several readily accessible methods were compared. The focus, however, was on demonstrating the optimization framework rather than performing exhaustive benchmarking or runtime tuning.

A two-stage optimization strategy \cite{schoenTwoPhaseMethodsGlobal2002b} was found to be the most reliable to locate the global minimum of the \RfactorDash{} hypersurface. In this approach, the optimization is split into an exploratory stage, which aims to identify the region containing the global minimum within the parameter bounds, and a local stage, which refines the structure in this region. The exploratory and local stages may employ different optimization algorithms tailored to their respective tasks.

\subsection{Exploratory Search Using CMA-ES}

The first, exploratory stage of the search has the task of finding the basin of the global minimum.
Simple parameter-wise optimization was found to be insufficient, as it is highly susceptible to trapping into local minima. Basin hopping \cite{walesGlobalOptimizationBasinHopping1997}, as implemented in \program{SciPy} \cite{2020SciPy-NMeth}, was also found to be inefficient, due to the computational cost of evaluating the gradient $\nabla R$ across large regions of the parameter space.

The covariance-matrix-adaptation evolution strategy (CMA-ES) is a robust optimization algorithm, particularly suitable for high-dimensional, nonlinear problems. It provides a balance between exploration and exploitation, is robust against local minima, and does not require evaluation of gradients \cite{hansen2001cma,hansenCMAEvolutionStrategy2023}.
Similar to the search algorithm in \tenserleed{} \cite{kottckeNewApproachAutomated1997}, it is based on generations of point clouds drawn from multivariate normal (Gaussian) distributions. Compared to the algorithm in \tenserleed, one of the advantages of CMA-ES is a sophisticated method for adapting the covariance matrix of the distribution to the \RfactorDash{} hypersurface.

The CMA-ES implementation from the \textsc{Clinamen2} package \cite{wanzenbockClinamen2FunctionalstyleEvolutionary2024} was employed due to its efficiency and compatibility with the \python{}- and JAX-based framework.
Because the standard CMA-ES implementation does not support parameter bounds, adaptations were applied to operate within a bounded parameter space. Details of this adaptation are provided in \SISection{11}.

\subsection{Switching from Exploratory to Local Optimization}

The two-stage approach requires a convergence criterion for switching from the the exploratory stage to local refinement.
While it is impossible to identify the global minimum with absolute certainty, it is easier to determine whether the exploratory optimization has converged on a single attraction basin.
Once this happens, the transition from the exploratory stage to the local one should be made.

To identify this basin is not trivial; several approaches are possible.
In this work, a threshold ($\approx{} 5 \times 10^{-3}$) was applied to the standard deviation of $R$ factors within a single CMA-ES generation. Once this threshold is reached, the optimization proceeds to the local stage.
This criterion is based on the assumption that when CMA-ES generates configurations with similar \RfactorDash{} values, the algorithm has likely identified the desired region. Alternatively, a criterion based on 
\RfactorDash{} changes between generations could be conceived.

Information obtained during the exploratory stage can be used to improve the local optimization. For example, the covariance matrix estimated in CMA-ES is proportional to the inverse of the Hessian of the \RfactorDash{} hypersurface \cite{shirCovarianceHessianRelationEvolution2020} and can serve to precondition the local refinement, as discussed \SISection{12}.

\subsection{\textit{R}-Factor Gradients for Local Refinement}

In optimization, the availability of gradients is known to improve optimization efficiency by indicating the direction of steepest descent.
Previous structure-optimization implementations in tensor LEED have not used analytical gradients of the \RfactorDash{} hypersurface. This limitation has been noted in the literature \cite{vanhoveLowEnergyElectronDiffraction1986,overOptimizationMethodsTheir1992,wanderNewModularLow2001}, and existing methods either rely on gradient-free algorithms or on finite-difference approximations \cite{rousDirectedSearchMethods1990}.

\begin{figure}
  \includegraphics[width=9.0cm]{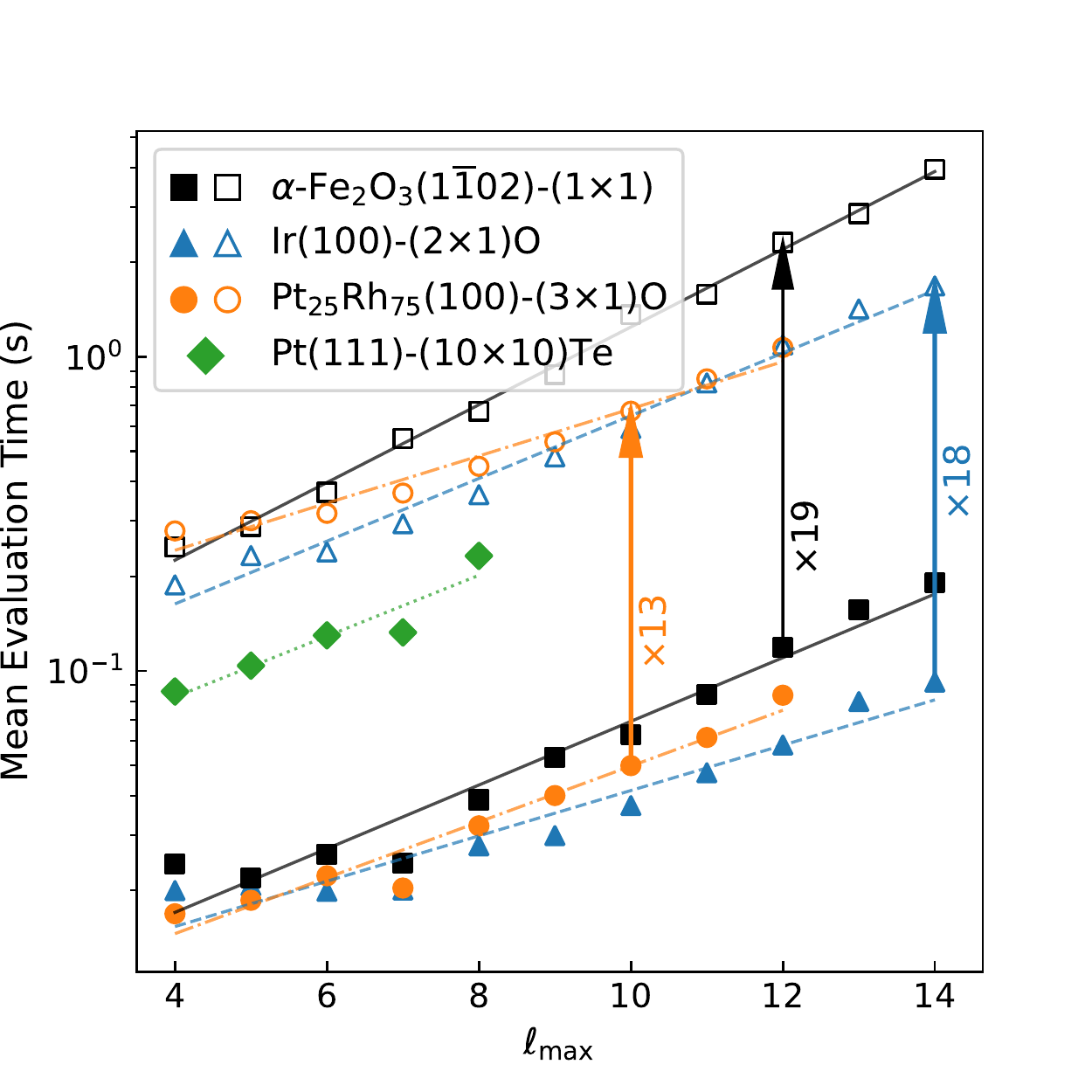}
  \caption{\label{fig:timing}
  Timing benchmarks for the calculation of \PendryR{} and its gradient for the test systems: \hematiteOneby{} (described in \Section{sec:benchmarks}), \iridiumox{}, and \ptrh{} (described in \SISection{14}). The number of free (irreducible) parameters for these systems are 33, 14 and 40 respectively.
  Timing benchmarks for the evaluation of \PendryR{} are also shown for \pt10x10{} (119 free parameters; no gradients calculated due to memory limitations, see \SISection{14} for details).
  The computing times (with an Nvidia A100 GPU) are shown as a function of the angular momentum quantum number cutoff $\ell_{\mathrm{max}}$.
  $R_{\mathrm{P}}$-factor evaluations are shown with solid symbols, while gradient evaluations are shown with open symbols.
  The evaluation time scales almost exponentially with $\ell_{\mathrm{max}}$.
  Dashed lines indicate the exponential fit of the timing data against the cutoff $\ell_{\mathrm{max}}$.
  Arrows indicate the factor by which the gradient calculation is slower than the \RfactorDash{} calculation at the respective \lmax{} value. }
\end{figure}

With the automatic-differentiation capabilities of JAX, the gradient $\nabla R$ of the \RfactorDash{} hypersurface with respect to the irreducible parameters $\tilde{\xi}$  can be calculated. Most functions in the tensor-LEED calculation are differentiable, but some cases require special treatment, such as the transformation between Cartesian and spherical coordinates \cite{abramowitz+stegun}. Details of these cases are provided in \SISection{4}.

The calculation of the \RfactorDash{} gradient is computationally more expensive than a simple evaluation of the \Rfactor, since gradients require the propagation of derivatives through all relevant functions.
\Fig{fig:timing} presents timing benchmarks for a few representative systems: the \hematiteOneby{} surface \cite{kraushoferViPErLEEDPackageCalculation2025} further discussed in  \Section{sec:hematite_oneby}, and the \iridiumox{} \cite{ferstlStructureOrderingOxygen2016}, \ptrh{} \cite{spornQuantitativeLEEDAnalysis1998}, and \pt10x10{} \cite{kisslingerSurfaceTelluridePhases2023} surfaces from \SISection{14}.
The mean evaluation times for the \Rfactor{} (solid symbols) and its gradient (open symbols) are plotted as a function of the cutoff of the angular-momentum quantum number \lmax{}.
Calculating the gradient increases the computing time by more than an order of magnitude.
On the other hand, the information content of the gradient is much larger than that of a single number ($n$ values for $n$ irreducible parameters), and numeric differentiation by finite differences ($n+1$ evaluations of the \Rfactor{}) would be slower than automatic differentiation for the larger systems.
Consequently, the choice to use gradients involves a tradeoff between higher computational cost and improved efficiency of each optimization step.
As discussed in \Section{sec:search_algorithms}, selective use of gradients is recommended.

\subsection{Local Refinement}
\label{sec:local_refinement}

Once a basin in the \RfactorDash{} hypersurface has been identified by the exploratory search, an algorithm for local optimization can be applied to refine the structure. For this purpose, several algorithms from \program{SciPy}'s \texttt{optimize.minimize} module \cite{2020SciPy-NMeth} were tested, including both gradient-based and gradient-free methods.

The choice of algorithm has a substantial impact on both success rate and runtime.
Algorithms suitable for medium- to high-dimensional optimization problems and robust against the small-scale noise \cite{rousTensorLEEDApproximationSurface1992,Imre2025PendryRfactor} in \PendryR{} (i.e., the objective function) are particularly advantageous.

\begin{figure}[tb!]
  \includegraphics[width=8.6cm]{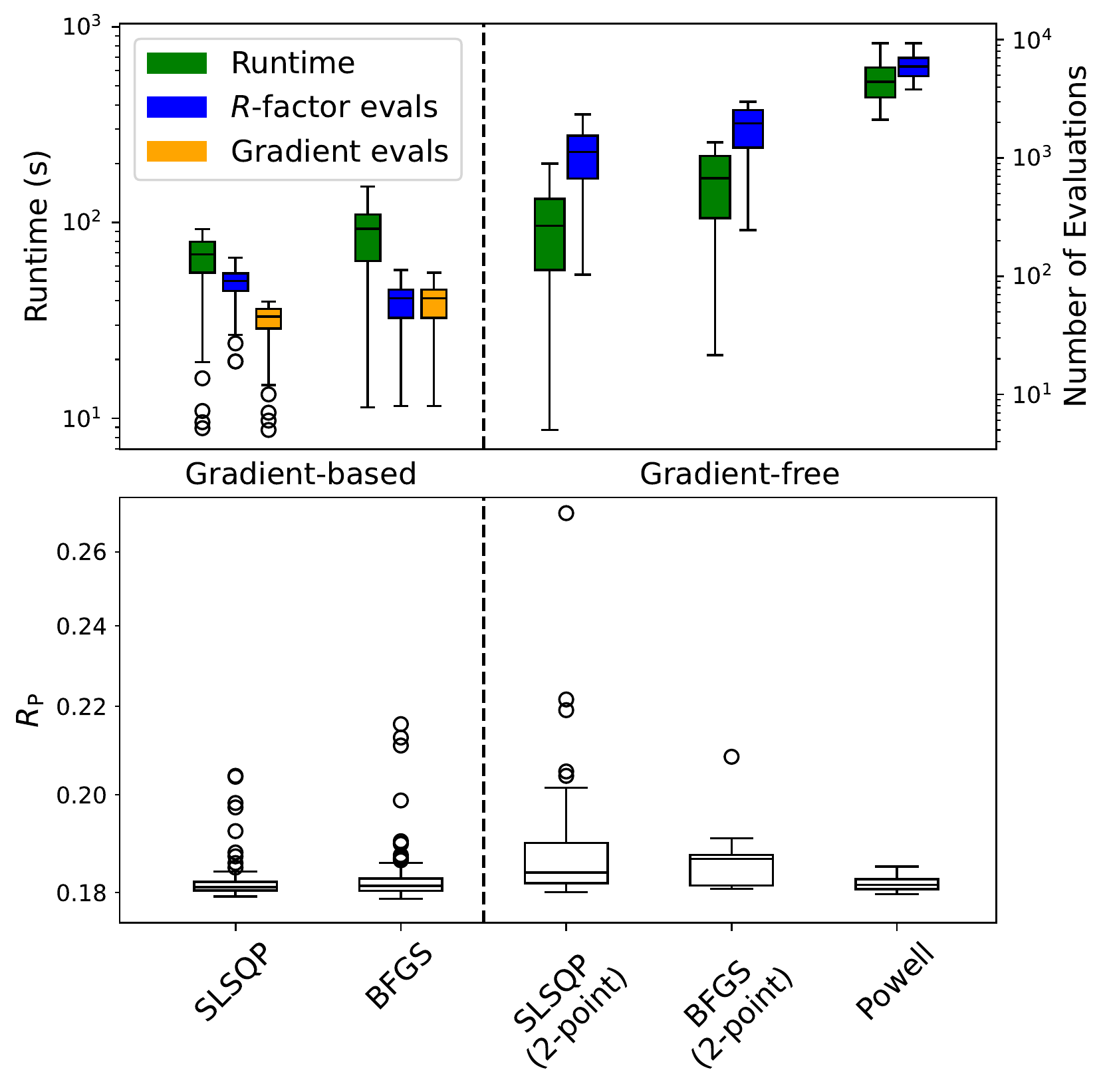}
  \caption{\label{fig:local_algos}
    Comparison of runtime and number of evaluations (top), as well as reliability (bottom) of local optimization algorithms near the global minimum.
    Algorithms were compared using the second segment (48 irreducible parameters) of the \hematiteOneby{} optimization described in \Section{sec:hematite_oneby}, starting from 98 configurations (one per generation) selected from the CMA-ES run.
    The algorithms shown were found to be the most reliable:
    sequential least-squares quadratic programming (SLSQP) \cite{kraftSoftwarePackageSequential1988}, Broyden--Fletcher--Goldfarb--Shanno (BFGS) \cite{nocedalNumericalOptimization2006}, and
    Powell \cite{powellEfficientMethodFinding1964,pressNumericalRecipesArt2007},
    as implemented in \program{SciPy} \cite{2020SciPy-NMeth}.
    Gradient-based algorithms are shown to the left of the dashed line, while finite-difference methods are shown to the right.
    Algorithms labeled as \enquote{2-point} employ two-point finite-difference approximations for the gradients.
    Computational details are provided in \SISection{3}. 
    In each box plot, boxes extend from the first to the third quartile; whiskers extend to the furthest data point within $1.5 \times$ of the inter-quartile range. Small circles indicate single outliers beyond the range of the whiskers.
    Tests using an improved \Rfactor{} \cite{Imre2025PendryRfactor} show that outliers in the bottom panel are due to the noisiness of \PendryR{}.
    }
\end{figure}

An overview of runtimes and \RfactorDash{} ranges for the most suitable algorithms is presented in \Fig{fig:local_algos}. Gradient-based algorithms (left) were more reliable---at comparable runtimes---than their finite-difference counterparts (right). Among these, the sequential least-squares quadratic programming (SLSQP) algorithm \cite{kraftSoftwarePackageSequential1988} and the quasi-Newton Broyden--Fletcher--Goldfarb--Shanno (BFGS) algorithm \cite{nocedalNumericalOptimization2006} were found to perform best. While BFGS outperformed SLSQP in certain cases, SLSQP is usually preferable, due to its less frequent use of computationally expensive gradient evaluations, as shown in \Fig{fig:local_algos} (orange).

When gradients are unavailable---for example, due to memory limitations---finite-difference versions of the algorithms provide a reasonable runtime--performance tradeoff. Powell’s algorithm \cite{powellEfficientMethodFinding1964,pressNumericalRecipesArt2007} offers a particularly reliable alternative (as indicated by the lack of outliers in the lower part of \Fig{fig:local_algos}), albeit at the cost of a substantially increased runtime (cf.\ top panel of \Fig{fig:local_algos}).

Additional algorithms implemented in \program{SciPy}, including COBYLA \cite{powellDirectSearchOptimization1994,powellDirectSearchAlgorithms1998,powellViewAlgorithmsOptimization2007}, TNC \cite{nashNewtonTypeMinimizationLanczos1984,nocedalNumericalOptimization2006}, conjugate gradient \cite{nocedalNumericalOptimization2006}, and Nelder--Mead \cite{nelderSimplexMethodFunction1965}, were also tested but found to be substantially inferior in terms of runtime, reliability, or both. The lower-memory variant L-BFGS-B \cite{byrdLimitedMemoryAlgorithm1995,zhuAlgorithm778LBFGSB1997}, often recommended for high-dimensional problems, did not outperform SLSQP or BFGS in these tests.

\section{Performance Benchmarks}
\label{sec:benchmarks}

To demonstrate the feasibility, and evaluate the performance of the presented tensor-LEED implementation, a series of benchmark tests on several known surface structures were performed.
\Section{sec:hematite_oneby} shows the results of a structural optimization of \hematiteOneby{}, which was used as a main example in the \calc{} publication \cite{kraushoferViPErLEEDPackageCalculation2025}.
Structure optimizations were also performed for the \iridiumox{} \cite{ferstlStructureOrderingOxygen2016}, \ptrh{} \cite{spornQuantitativeLEEDAnalysis1998}, and \pt10x10{} structures \cite{kisslingerSurfaceTelluridePhases2023}, which are discussed in \SISection{14}.
\cite{SupplementaryDatasetStructural}

\subsection{\boldmath{\hematiteOneby{}}\label{sec:hematite_oneby}}

Structural optimization of a bulk-truncated \hematiteOneby{} surface was performed using previously determined values for the incidence angles and the imaginary part of the inner potential \cite{kraushoferViPErLEEDPackageCalculation2025}. Optimization employed both \vprjax{} and the \tenserleed{} backend in \calc{}.
A reference calculation of the initial bulk-truncated structure gives $R_{\mathrm{P}}=0.747$.

Geometric parameters and vibration amplitudes were optimized in three segments of alternating reference calculations and tensor-LEED--based optimizations. Between these segments, intermediate reference calculations were used to reduce the errors introduced by the tensor-LEED approximation. The parameter ranges were kept consistent between implementations. In the first segment, all atoms in the top two layers were allowed to move in all symmetry-allowed directions, and the vibration amplitudes of the topmost Fe and O atoms were optimized.
This corresponds to 30 geometric and 2 vibrational parameters, plus the shift of the real part of the inner potential, for a total of 33 irreducible parameters.
In the second and third segments, also the atoms in the third layer were allowed to relax in all symmetry-allowed directions, adding 15 geometric parameters for a total of 48 irreducible parameters.

Both implementations achieve similar Pendry \RfactorDash{} values after the third segment: $R_{\mathrm{P}}=0.163$ (\tenserleed{}) and $R_{\mathrm{P}}=0.165$ (\vprjax{}). 
Further refinement is possible with both implementations, but does not aid the discussion here.
All differences in geometric and vibration amplitudes are well within the associated fit uncertainties, indicating that both implementations converge towards the same minimum configuration.
Minor differences in the final \Rfactor{} can be attributed to two main sources:
Firstly, \vprjax{} prioritizes numerical accuracy over exact reproduction of the \tenserleed{} results. Different implementations of the underlying routines (e.g., interpolation of diffraction intensities) can affect the exact \RfactorDash{} value.
Secondly, as mentioned in \Section{sec:local_refinement},  \PendryR{} forms a rough hypersurface with many local minima in the basin \cite{Imre2025PendryRfactor}. This means that even when using suitable optimization algorithms, it is impossible to gauge whether the optimization result corresponds to the global minimum. This problem disappears when using an improved \Rfactor{} \cite{Imre2025PendryRfactor}.

\begin{figure}[tb!]
  \includegraphics[width=8.5cm]{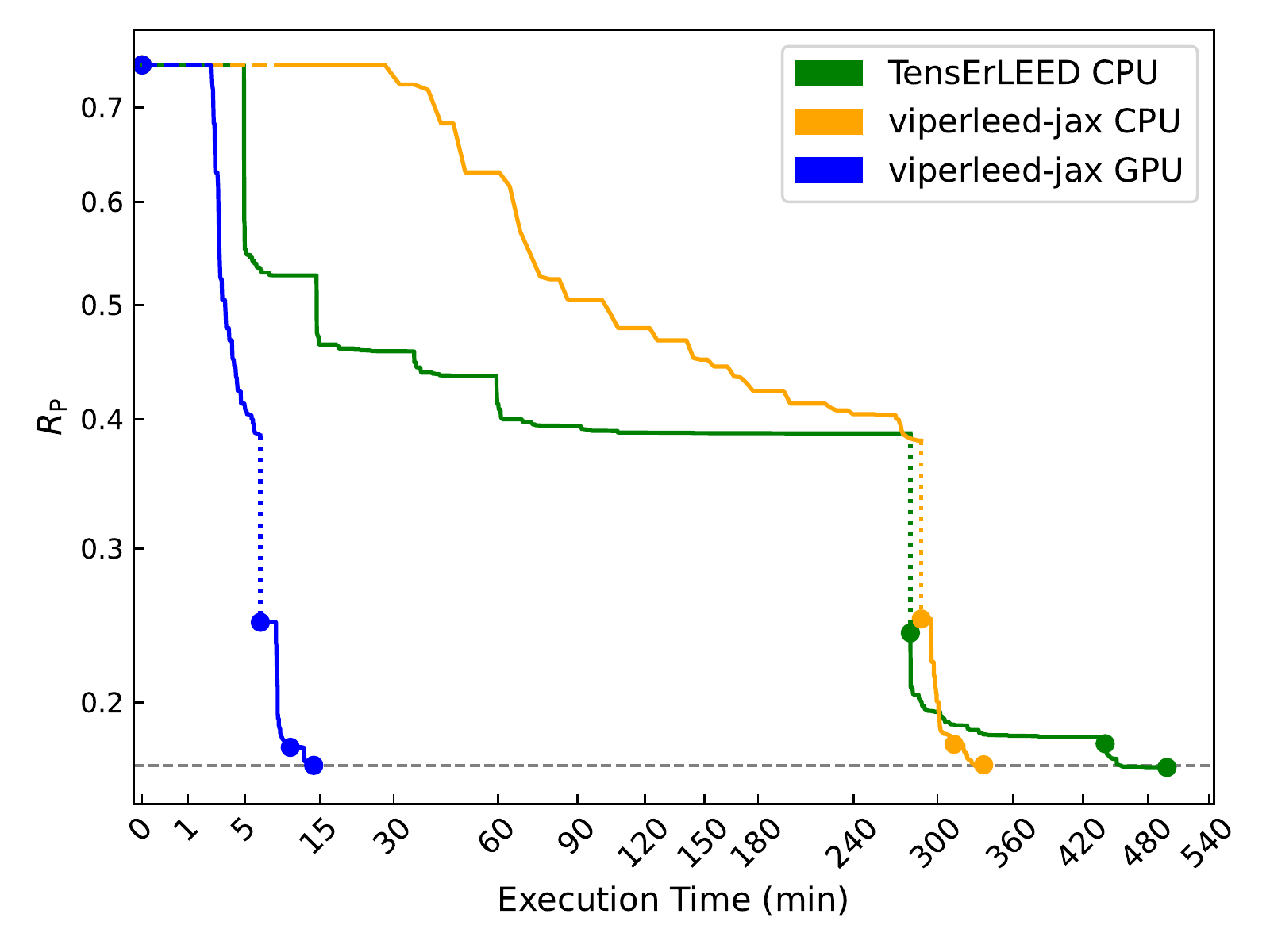}
  \caption{\label{fig:convergence}%
  Comparison of structure optimization for the \hematiteOneby{} surface using \vprjax{} and the \tenserleed{} backend in \calc{}.
  The running minimum of the \Rfactor{} is shown as a function of execution time.
  Optimization proceeds in three segments of alternating reference calculations and tensor-LEED-based optimizations.
  The starting configuration for each segment corresponds to the final structure of the previous segment, with the first segment beginning from a bulk-truncated structure.
  \RfactorDash{} values of the reference calculations are shown as filled circles.
  Broken horizontal lines indicate the time required for setup and just-in-time compilation in the new tensor-LEED implementation.
  Dotted vertical lines mark differences in the \Rfactor{} values between the end of an optimization segment and the subsequent reference calculation, reflecting errors from the perturbative tensor-LEED approximation that are corrected by the reference calculations.
  A dashed gray line indicates the smallest \RfactorDash{} value ($R_{\mathrm{P}}=0.165$) achieved by the new tensor-LEED implementation.
  }
\end{figure}

The optimization progress is shown in \Fig{fig:convergence} as \Rfactor{} versus execution time.
Times for the reference calculations (170\,s in each segment) are excluded, as they are independent of the tensor-LEED implementation.
To allow for an unbiased comparison, the computing times in \Fig{fig:convergence} include setup and just-in-time compilation for \vprjax{}, and the \deltas{} calculation for the \tenserleed{} backend.

On a computing node with 128 CPU cores (AMD 7713) of a modern cluster, \calc{} with the \tenserleed{} backend requires approximately 8.5~h for all three segments (green in \Fig{fig:convergence}).
With the same hardware, \vprjax{} completes in less than 5.5~h (orange in \Fig{fig:convergence}).
While JAX is not primarily optimized for CPU execution, it still uses just-in-time compilation and thread-based parallelization on CPUs.
Running the same calculation on a single Nvidia A100 GPU reduces the total runtime to less than 15~min (blue in \Fig{fig:convergence}).

\begin{figure}[h!]
  \includegraphics[width=7.7cm]{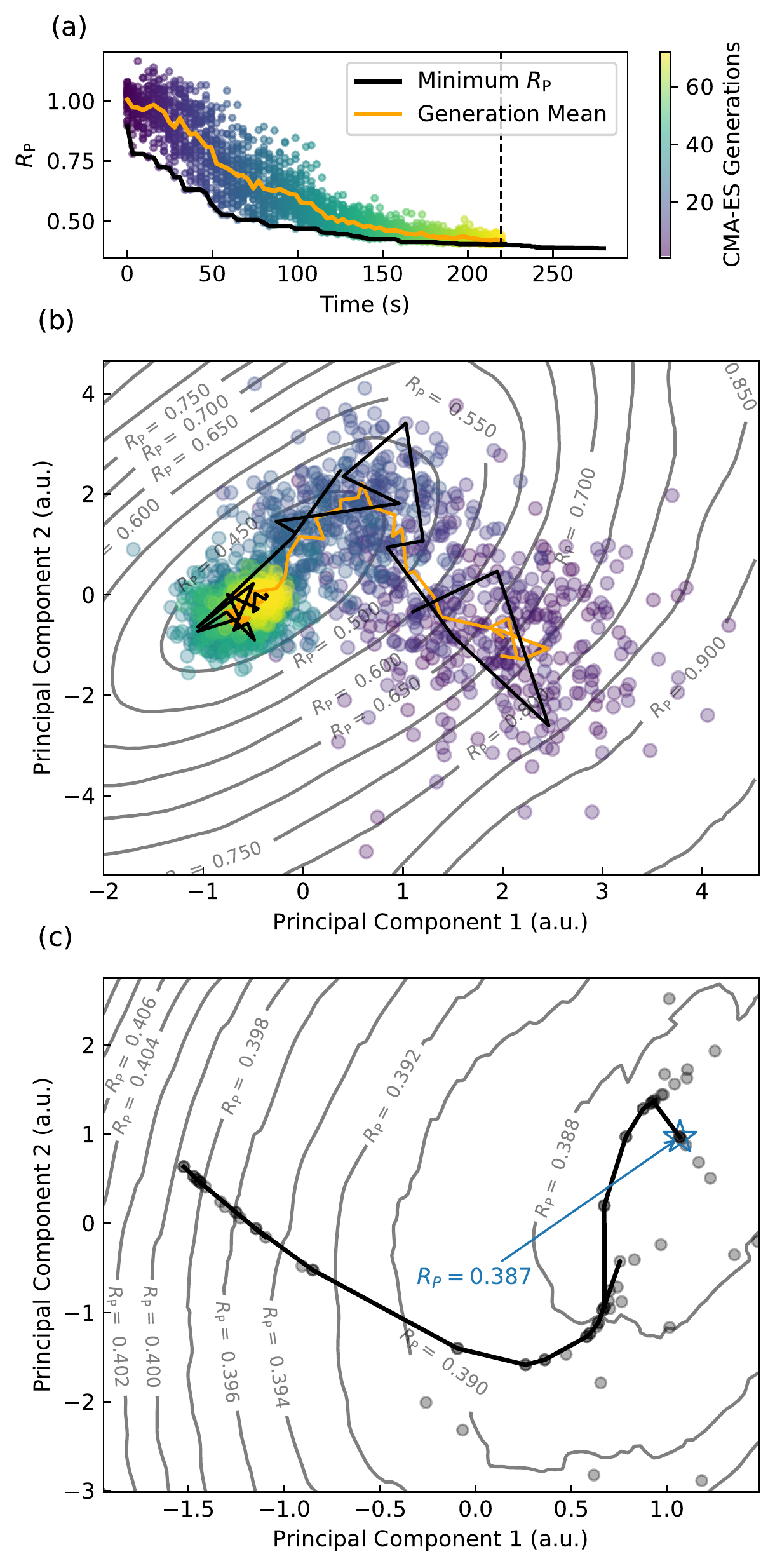}
  \caption{\label{fig:trajectory}
  Structure optimization for \hematiteOneby{} using a two-stage minimization approach. Exploratory and local stages used CMA-ES \cite{wanzenbockClinamen2FunctionalstyleEvolutionary2024} and SLSQP \cite{2020SciPy-NMeth} algorithms, respectively. (a) $R_{\mathrm{P}}$ factor as a function of time. The dashed line marks the switch from the exploratory to the local optimization stage. (b) Points evaluated by the CMA-ES algorithm, projected onto two dimensions using PCA. Colors indicate generation numbers with the same scheme as in panel (a). Contour lines display the \RfactorDash{} in a plane spanned by the first two principal components.
  The mean and the individual with the smallest \Rfactor{} for each generation are highlighted by orange and a black lines, respectively.
(c) Trajectory of the local optimization, projected onto a 2D plane [different from that of panel (b)] to highlight convergence towards the minimum. Evaluated configurations and the running minimum are represented as individual points and as a black line, respectively.
The blue star marks the final configuration with $R_{\mathrm{P}}=0.387$.
  }
\end{figure}

\subsection{Discussion of the Optimization Process}
\label{sec:discussion_optimization_process}

Using the \hematiteOneby{} example, it is worth taking a closer look at the optimization process.
The first optimization segment involves 33 irreducible parameters, automatically identified and analyzed by the parameter-tree algorithm based on the user input.

The first segment employs the two-stage optimization approach described in \Section{sec:search_algorithms}.
In the exploratory stage, CMA-ES searches for the basin containing the global minimum of the \RfactorDash{} hypersurface.
Using generations of 30 individuals, CMA-ES initially explores the parameter space in the wide range given by the parameter bounds, homing in on the region of the global minimum within fewer than 100 generations, corresponding to a runtime of less than 7~min (including setup and just-in-time compilation) on a single Nvidia A100 GPU.
This process is illustrated in \Fig{fig:trajectory}, where \SubFig{fig:trajectory}{(a)} shows the \Rfactor{} as a function of optimization time, and \SubFig{fig:trajectory}{(b)} projects the CMA-ES trajectory onto a two-dimensional subspace of the parameter space obtained by principal-component analysis (PCA). When viewing this plot type, the reader should be aware that the contour lines show the \Rfactor{} in a plane though the minimum, but there are 31 dimensions orthogonal to the plane plotted. The plane plotted is that with the largest influence on the \Rfactor. In other words, the points sampled (with the exception of the final one) are not exactly in the plane with the contour lines. Thus, the \RfactorDash{} values along the trajectory  in \SubFig{fig:trajectory}{(a)} differ from the values in the contour plot. The \Rfactor{} continuously decreases along the black line in \SubFig{fig:trajectory}{(a)} and (b), but this is not seen in the 2D plot of \SubFig{fig:trajectory}{(b)}, since the 2D projection cannot show all the out-of-plane dimensions. Nevertheless, the plot illustrates that the mean of the CMA-ES population (orange line) approaches the region of the global minimum (yellow).

After the region around the global minimum is located, the optimization switches to the local stage, using the SLSQP algorithm \cite{kraftSoftwarePackageSequential1988} to refine the structure. This decreases the \Rfactor{} from $R_{\mathrm{P}}=0.403$ to $0.387$ by the end of the first segment. This corresponds to the minimum of the \RfactorDash{} hypersurface as calculated in the tensor-LEED approximation based on a perturbation of the initial bulk-truncated surface. The computing time for the local-minimum search is 60\,s.

\SubFig{fig:trajectory}{(c)} illustrates this stage, where the trajectory is again projected onto a two-dimensional PCA subspace [different from that in \SubFig{fig:trajectory}{(b)}; again the remaining 31 dimensions cannot be shown]. The contour lines also show the roughness (noisiness) of Pendry's \Rfactor, which may lead to a partially erratic trajectory; a problem that can be avoided by an improved \Rfactor{} \cite{Imre2025PendryRfactor}.

The second and third segments proceed similarly but omit the exploratory optimization stage, relying solely on gradient-based refinement. Detailed convergence information for these segments is provided in \SISection{13}.

It is worth pointing out that the performance improvements compared with the \tenserleed{} search is due to a huge reduction of the number of \RfactorDash{} evaluations---each including calculation of \iv{} curves--- rather than just a reduction in the number of $\delta A_{\mathbf{g}}$ calculations.
\vprjax{} and \tenserleed{} require a comparable number of $\delta A_{\mathbf{g}}$ evaluations, but the on-the-fly calculation in \vprjax{} results in a higher overall efficiency
\footnote{
    In \vprjax{}, approximately $2.3 \times 10^3$ \RfactorDash{} values were evaluated in the first segment of the \hematiteOneby{} example.  In contrast, the algorithm of \citet{kottckeNewApproachAutomated1997}, as used in \tenserleed{}, requires about $5.7\times10^5$ generations with 128 individuals each to reach the same \RfactorDash{} value, corresponding to roughly $7\times10^7$ \RfactorDash{} evaluations.

    In \vprjax{}, \RfactorDash{} evaluations are directly coupled with the \deltas{} calculations, resulting in approximately $2.3 \times 10^3$ \deltas{} evaluations per atom under variation, yielding about $7\times10^4$ evaluations for the 30 atoms in the first stage.

    In \tenserleed{}, $\delta A_{\mathbf{g}}$ values are only evaluated during the \enquote{delta-amplitude calculation}, where they are precomputed on a user-defined grid. In the first segment of the \hematiteOneby{} example, a grid of 11, 7, and 7 geometric steps in $x$, $y$, and $z$ directions, respectively, was used for each of the 30 perturbed atoms. The four topmost atoms were additionally allowed 9 different vibration-amplitude values, which were varied together with the $z$ coordinate. Altogether, this yields $26\times(11+7+7) + 4\times(11\times9+7+7) = 1102$ atomic \deltas{} evaluations per search loop, and around $17\times 10^4$ evaluations for the 16 loops required for convergence.
    }.

\section{Conclusions}

This work introduces a generalized approach to structure optimization in \leediv{}, extending the \calc{} package. A tree-like representation of the parameter space reduces the problem to an irreducible set of normalized variables, decoupling the optimization from the details of the tensor-LEED technique. This enables the direct use of established bounded optimization algorithms and, compared with earlier \leedivdash{} implementations, leverages decades of development in numerical methods, yielding more robust and efficient refinements.

Central to this work is a fully JAX-based tensor-LEED implementation. Developed entirely in \python{}, it provides differentiable \RfactorDash{} calculations and enables gradient-based optimization, marking the first use of non-finite-difference gradients in \leedivdash{} structure determination via tensor LEED.

A two-stage optimization strategy proved particularly effective: CMA-ES for exploration, followed by gradient-based local refinement. This combination consistently identifies the global minimum of the \RfactorDash{} hypersurface and outperforms the \tenserleed{}-based search in \calc{}. As the \RfactorDash{} evaluation now is a black-box objective function, the new tensor-LEED implementation developed in this work provides the ground for future tests of promising optimization algorithms beyond those examined here, such as particle-swarm optimization \cite{duncanGlobalSearchAlgorithms2012} and machine-learning–inspired methods such as Bayesian optimization or stochastic gradient descent \cite{kochenderferAlgorithmsOptimization2019} on subsets of beams. 

For comparability with previous work, the test cases presented here use Pendry's \Rfactor{}, which is a noisy objective function for minimization (a \enquote{rough} \RfactorDash{} surface). In terms of both performance and reliability, the gradient-based local refinement will greatly benefit from an improved \Rfactor{} already implemented in \vprjax, eliminating the roughness and multitude of local \RfactorDash{} minima in the basin of the global minimum \cite{Imre2025PendryRfactor}.

Although developed for tensor LEED, the principles are general and may also be valuable for other methods like surface x-ray diffraction. The modular code design supports future extensions, such as additional constraints or nonstructural parameters.

\section{Data and Code Availability}
The source code for \vprjax{} is available online in \RefInline{imreViperleedjax} under the GNU General Public License version 3 (or later) \cite{gnuv3}.
The raw data used for the examples presented in this work are available at \cite{SupplementaryDatasetStructural}.

\section{Author contributions}

A.M.I. conceived the presented formalism. A.M.I. and P.H. developed the tensor-LEED software implementation and structure-optimization routines.
The technical implementation was aided by F.K., R.W., F.B., T.H., and M.R..
S.T., M.K., J.C., G.K.H.M., M.S., and U.D. supported development of the main ideas and data analysis.
M.S., U.D., and M.R. supervised the project.
The manuscript was drafted by A.M.I. with input from all authors.

\section{Competing interests}
The authors declare no competing interests.

\begin{acknowledgments}
The authors would like to thank Lutz Hammer for carefully reading the manuscript and many helpful suggestions. 
This research was funded in part by the Austrian Science Fund (FWF) under \doiRef{10.55776/F8100}, Taming Complexity in
Materials Modeling (TACO), and under \doiRef{10.55776/J4811} (F.K.), as well as by the European Research Council (ERC) under the European Union’s Horizon 2020 research and innovation programme (grant agreement No. 883395, Advanced Research Grant ‘WatFun’).
J.C. acknowledges grant CEX2023-001286-S funded by MICIU/AEI \doiRef{10.13039/501100011033} and grant PID2023-148359NB-C21 funded by MICIU/AEI \doiRef{10.13039/501100011033} and the European Union FEDER.
The computational results presented have been achieved in part using the Austrian Scientific Computing (ASC) infrastructure.
For the purpose of open access, the authors have applied a CC BY public copyright license to any Author Accepted Manuscript version arising from this submission.

\end{acknowledgments}

\providecommand{\enquote}[1]{#1}
\bibliography{tensor-leed}

\end{document}